\documentclass[twocolumn,nolinenumbers]{aastex631}
\usepackage{amsmath}
\usepackage{multirow}

\begin{document}

\title{X-ray reflection spectroscopy with improved calculations of the emission angle}

\author[0009-0008-3299-9185]{Yimin~Huang}
\affiliation{Center for Astronomy and Astrophysics, Center for Field Theory and Particle Physics, and Department of Physics,\\
Fudan University, Shanghai 200438, China}

\author[0000-0003-2845-1009]{Honghui~Liu}
\altaffiliation{honghui.liu@uni-tuebingen.de}
\affiliation{Institut f\"ur Astronomie und Astrophysik, Eberhard-Karls Universit\"at T\"ubingen, D-72076 T\"ubingen, Germany}

\author[0000-0002-9337-4612]{Temurbek~Mirzaev}
\affiliation{Center for Astronomy and Astrophysics, Center for Field Theory and Particle Physics, and Department of Physics,\\
Fudan University, Shanghai 200438, China}

\author[0009-0009-2549-1161]{Ningyue~Fan}
\affiliation{Center for Astronomy and Astrophysics, Center for Field Theory and Particle Physics, and Department of Physics,\\
Fudan University, Shanghai 200438, China}

\author[0000-0002-3180-9502]{Cosimo~Bambi}
\altaffiliation{bambi@fudan.edu.cn}
\affiliation{Center for Astronomy and Astrophysics, Center for Field Theory and Particle Physics, and Department of Physics,\\
Fudan University, Shanghai 200438, China}
\affiliation{School of Natural Sciences and Humanities, New Uzbekistan University, Tashkent 100007, Uzbekistan}

\author[0000-0003-0847-1299]{Zuobin~Zhang}
\affiliation{Astrophysics, Department of Physics, University of Oxford, Oxford OX1 3RH, UK}

\author[0000-0003-4583-9048]{Thomas~Dauser}
\affiliation{Dr. Karl Remeis-Observatory and Erlangen Centre for Astroparticle Physics, D-96049 Bamberg, Germany}

\author[0000-0003-3828-2448]{Javier~A.~Garc{\'\i}a}
\affiliation{NASA Goddard Space Flight Center, Greenbelt, MD 20771, USA}
\affiliation{Cahill Center for Astronomy and Astrophysics, California Institute of Technology, Pasadena, CA 91125, USA}

\author[0000-0002-5311-9078]{Adam Ingram}
\affiliation{School of Mathematics, Statistics, and Physics, Newcastle University, Newcastle upon Tyne NE1 7RU, UK}

\author[0000-0002-9639-4352]{Jiachen Jiang}
\affiliation{Department of Physics, University of Warwick, Coventry CV4 7AL, UK}

\author[0000-0003-4216-7936]{Guglielmo~Mastroserio}
\affiliation{Dipartimento di Fisica, Universit\`a degli Studi di Milano, I-20133 Milano, Italy}
\affiliation{Scuola Universitaria Superiore IUSS Pavia, Palazzo del Broletto, I-27100 Pavia, Italy}

\author{Shafqat~Riaz}
\affiliation{Institut f\"ur Astronomie und Astrophysik, Eberhard-Karls Universit\"at T\"ubingen, D-72076 T\"ubingen, Germany}

\author[0000-0003-3402-7212]{Swarnim~Shashank}
\affiliation{Center for Astronomy and Astrophysics, Center for Field Theory and Particle Physics, and Department of Physics,\\
Fudan University, Shanghai 200438, China}

\begin{abstract}
The reflection spectrum produced by a cold medium illuminated by X-ray photons is not isotropic and its shape depends on the emission angle. In the reflection spectrum of an accretion disk of a black hole, the value of the emission angle changes over the disk and, in general, is different from the value of the inclination angle of the disk because of the light bending in the strong gravitational field of the black hole. Current reflection models make some approximations, as calculating a reflection spectrum taking the correct emission angle at every point of the disk into account would be too time-consuming and make the model too slow to analyze observations. In a recent paper, we showed that these approximations are unsuitable to fit high-quality black hole spectra expected from the next generation of X-ray missions. Here, we present a reflection model with improved calculations of the emission angle that solves this problem.     
\end{abstract}

\section{Introduction} \label{sec:intro}

The accretion of matter onto black holes is crucial for studying the physics and astrophysics of these objects. The electromagnetic spectra of accreting black holes often contain extensive information about the accretion flow and the spacetime geometry~\citep{Reynolds_2013,Bambi_2017,book}, featuring both thermal and non-thermal components~\citep{Remillard_2006}.

Over recent decades, the analysis of X-ray spectra of black hole X-ray binaries and active galactic nuclei has revealed the existence of a reflection component, produced when a hot corona illuminates a cold accretion disk. Such a reflection component takes significant features, such as iron K$\alpha$ lines at 6.4-7~keV and a Compton hump at 20-50~keV. These elements are invaluable tools for studying the properties of accreting black holes. \citet{1989MNRAS.238..729F} were the first to compute relativistically broadened line profiles. The first clear detection of a broadened iron line was observed in the X-ray spectrum of MCG--06--30--15 \citep{1995Natur.375..659T}. The analysis of the shape of these broadened iron lines is a powerful tool to measure black holes spins~\citep{2021SSRv..217...65B,2024ApJ...969...40D} and test Einstein's theory of General Relativity in the strong field regime~\citep{2017ApJ...842...76B,2018PhRvL.120e1101C,2019ApJ...875...56T,2021ApJ...913...79T}.

The reflection spectrum is not emitted isotropically and its shape depends on the emission angle, $\theta_e$, which is the angle between the normal to the surface of the accretion disk and the propagation direction of the radiation \citep{Garcia_2014}. The inclination angle of the accretion disk, $\theta_{obs}$, refers instead to the angle between the normal to the accretion disk and the line of sight of the distant observer. In the reflection spectrum of an accretion disk of a black hole, the value of the emission angle $\theta_e$ changes over the disk and, in general, is different from the inclination angle $\theta_{obs}$ because of the effect of light bending in the strong gravitational field of the black hole.

The old generation of reflection models were designed to calculate relativistic reflection spectra by applying relativistic convolution models [e.g., {\tt kyrline}~\citep{2004ApJS..153..205D}, {\tt kerrdisk}~\citep{2006ApJ...652.1028B}, or {\tt relconv}~\citep{2010MNRAS.409.1534D}] to non-relativistic reflection spectra calculated for $\theta_e = \theta_{obs}$ [e.g., calculated by {\tt xillver}~\citep{Garc_a_2010}]. By design, a convolution model is unsuitable to account for the distribution of different emission angles seen by an observer with a certain viewing angle.

\citet{Garcia_2014} were the first, with the model {\tt relxill}, to combine a relativistic convolution model and a non-relativistic reflection model to take the difference between $\theta_e$ and $\theta_{obs}$ into account. The impact of the assumption $\theta_e = \theta_{obs}$ on the estimate of the parameters of a source depends in a non-trivial way on the black hole spin parameter, disk ionization, and inclination angle $\theta_{obs}$, as well as on the characteristics of the X-ray detector and the quality of the data. In some cases, the assumption $\theta_e = \theta_{obs}$ works quite well, but in other cases there are non-negligible systematic effects in the estimate of some parameters \citep[see, e.g.,][]{Garcia_2014,2016MNRAS.457.1568M,2019MNRAS.488..324I,2020MNRAS.498.3565T}.

Even current reflection models like {\tt relxill} do not calculate relativistic reflection spectra by using the non-relativistic reflection spectra calculated with the correct emission angle at every point of the disk, as the calculations would be too time-consuming and the models would become too slow to analyze observations. In the case of {\tt relxill}, the model first calculates an average non-relativistic reflection spectrum, $\langle I_e \rangle$, and then uses $\langle I_e \rangle$ for the whole accretion disk. In the case of the lamppost model of the {\tt relxill} package, {\tt relxilllp}, the model divides the accretion disk into 10~annuli, calculates the average non-relativistic reflection spectrum for every annulus, then calculates the relativistic reflection spectrum for every annulus, and eventually sums the reflection spectra of all annuli. In the case of {\tt relxill\_nk}~\citep{Abdikamalov_2019,2020ApJ...899...80A}, which is an extension to non-Kerr spacetimes of the {\tt relxill} package, the model divides the accretion disk into 50~annuli in order to improve the accuracy of the final result. However, the value of the emission angle depends on the azimuth angle (especially close to the black hole and when the inclination angle of the disk is high), so dividing the accretion disk into annuli is still a rough approximation.

In \citet{2025MNRAS.536.2594L}, we used a ray-tracing code that can calculate relativistic reflection spectra by using the correct emission angle at every point of the accretion disk (and it is thus very accurate but too slow to be able to analyze observations) to simulate the observations of bright black hole X-ray binaries. We showed that even {\tt relxill} and {\tt relxill\_nk} cannot fit well high-quality black hole spectra expected from the next generation of X-ray missions and that this is due to the crude approximation of the average non-relativistic reflection spectrum $\langle I_e \rangle$ over the disk or the annuli.

Here we present an extension of the {\tt relxill} model with improved calculations of the emission angle. We call this model {\tt relxillA}, where A stands for ``angle improved''. The key-feature of {\tt relxillA} is that we divide the accretion disk in up to 10~zones. These zones are not annuli, as in the case of {\tt relxilllp} and {\tt relxill\_nk}. The cosine of the emission angles across the disk, $\mu = \cos\theta_{e}$, is divided into 10~equal-width intervals, ranging from 0 to 1 in steps of 0.1. Each interval represents a successive range of emission angles, ensuring a uniform sampling across the full angular distribution. Depending on the value of $\theta_{obs}$ and the black hole spin parameter $\alpha$, some accretion disks may not have all 10~zones. The relativistic reflection spectra calculated by the new model turn out to be much more accurate than those calculated by the existing models, and the new model is only moderately slower than the existing ones.

The paper is organized as follows. In Section~\ref{sec:2}, we review the calculations of a reflection spectrum in {\tt relxill}, {\tt relxilllp}, and {\tt relxill\_nk}. In Section~\ref{sec:2bis}, we present our angle-improved model \texttt{relxillA} and compare its predictions with an accurate but time-consuming ray-tracing code and with \texttt{relxill}. In Section~\ref{sec:sim}, we show that {\tt relxillA} can fit well the high-quality black hole spectra expected from the next generation of X-ray missions simulated in \citet{2025MNRAS.536.2594L}. In Section~\ref{sec:3}, we analyze a high-quality \textit{NuSTAR} spectrum of the black hole binary EXO~1846--031 with \texttt{relxill} and \texttt{relxillA} to evaluate the impact of the improved calculations of the emission angle on the estimate on the properties of an accreting black hole with current X-ray data. Summary and conclusions are reported in Section~\ref{sec:4}.

\section{Calculations in current reflection models} \label{sec:2}

The logical way to calculate the relativistic reflection spectrum of a specific system would follow three steps: $i)$ the calculation of how the hot corona illuminates the accretion disk, which requires to solve photon trajectories from the corona to the accretion disk, $ii)$ the calculation of the non-relativistic reflection spectrum at every radial coordinate of the accretion disk (assuming that the system is axisymmetric), which requires to solve certain radiative transfer equations, and $iii)$ the calculation of the accretion disk redshift image of the distant observer, which requires to solve photon trajectories backward in time, from the plane of the distant observer to the emission point on the accretion disk. At the end of these calculations, we have the accretion disk image as detected by the distant observer. Integrating over the image, we get the relativistic reflection spectrum of the accretion disk; see, e.g., \citet{bambi2024blackholexrayspectra} for a pedagogical and detailed presentation of these calculations.

These calculations are too time-consuming to be done during the data analysis process and therefore current reflection models follow a different approach. We can write the observed spectral flux as the following integral \citep[for more details, see, e.g.,][]{bambi2024blackholexrayspectra}
\begin{widetext}
\begin{eqnarray}\label{eq-flux}
F_{obs} (E_{obs}) &=& \frac{1}{D^2} \int_{R_{\rm in}}^{R_{\rm out}} dr_e \int_0^1 dg^* 
\frac{\pi r_e g^2}{\sqrt{g^* \left(1 - g^*\right)}} 
f^{(1)} (g^* , r_e , \theta_{obs}) \, I_e (E_e , r_e , \theta_e^{(1)}) \nonumber\\
&& + \frac{1}{D^2} \int_{R_{\rm in}}^{R_{\rm out}} dr_e \int_0^1 dg^* 
\frac{\pi r_e g^2}{\sqrt{g^* \left(1 - g^*\right)}} 
f^{(2)} (g^* , r_e , \theta_{obs}) \, I_e (E_e , r_e , \theta_e^{(2)}) \, ,
\end{eqnarray}
\end{widetext}
where $E_{obs}$ and $E_e$ are, respectively, the photon energy in the rest-frame of the distant observer and of the material in the disk; $D$ is the distance between the observer and the black hole; $r_e$ is the emission radius; $R_{\rm in}$ and $R_{\rm out}$ are, respectively, the inner and the outer radius of the accretion disk; $g = E_{obs}/E_e$ is the redshift factor and $g^*$ ($0 \le g^* \le 1$) is the relative redshift factor
\begin{eqnarray}
g^* = \frac{g - g_{\rm min}}{g_{\rm max} - g_{\rm min}} \, ,
\end{eqnarray}
where $g_{\rm min} = g_{\rm min} (r_e , \theta_{obs})$ and $g_{\rm max} = g_{\rm max} (r_e , \theta_{obs})$ are, respectively, the minimum and maximum values of the redshift factor at the emission radius $r_e$ for an observer with inclination angle $\theta_{obs}$ (for a specific spacetime metric). $f$ is the Cunningham's transfer function~\citep{1973ApJ...183..237C}
\begin{eqnarray}
f^{(i)} = \frac{g \sqrt{g^* \left( 1 - g^* \right)}}{\pi r_e} \left|\frac{\partial \left(X,Y\right)}{\partial \left(r_e,g^*\right)}\right|\, ,
\end{eqnarray}
where $X$ and $Y$ are the Cartesian coordinates of the image plane of the distant observer and $\partial(X,Y)/\partial(r_e,g^*)$ is the Jacobian between the coordinates $(X,Y)$ used on the image plane of the distant observer and the coordinates $(r_e,g^*)$ used on the disk. In Eq.~(\ref{eq-flux}) we have two transfer functions, $f^{(1)}$ and $f^{(2)}$, because at every emission radius we have two branches connecting the emission point of $g_{\rm min}$ with that of $g_{\rm max}$. $I_e$ is the specific intensity of the radiation at the emission point and depends on the emission angle $\theta_e$.

\begin{figure*}
    \centering
    \includegraphics[width=1\linewidth]{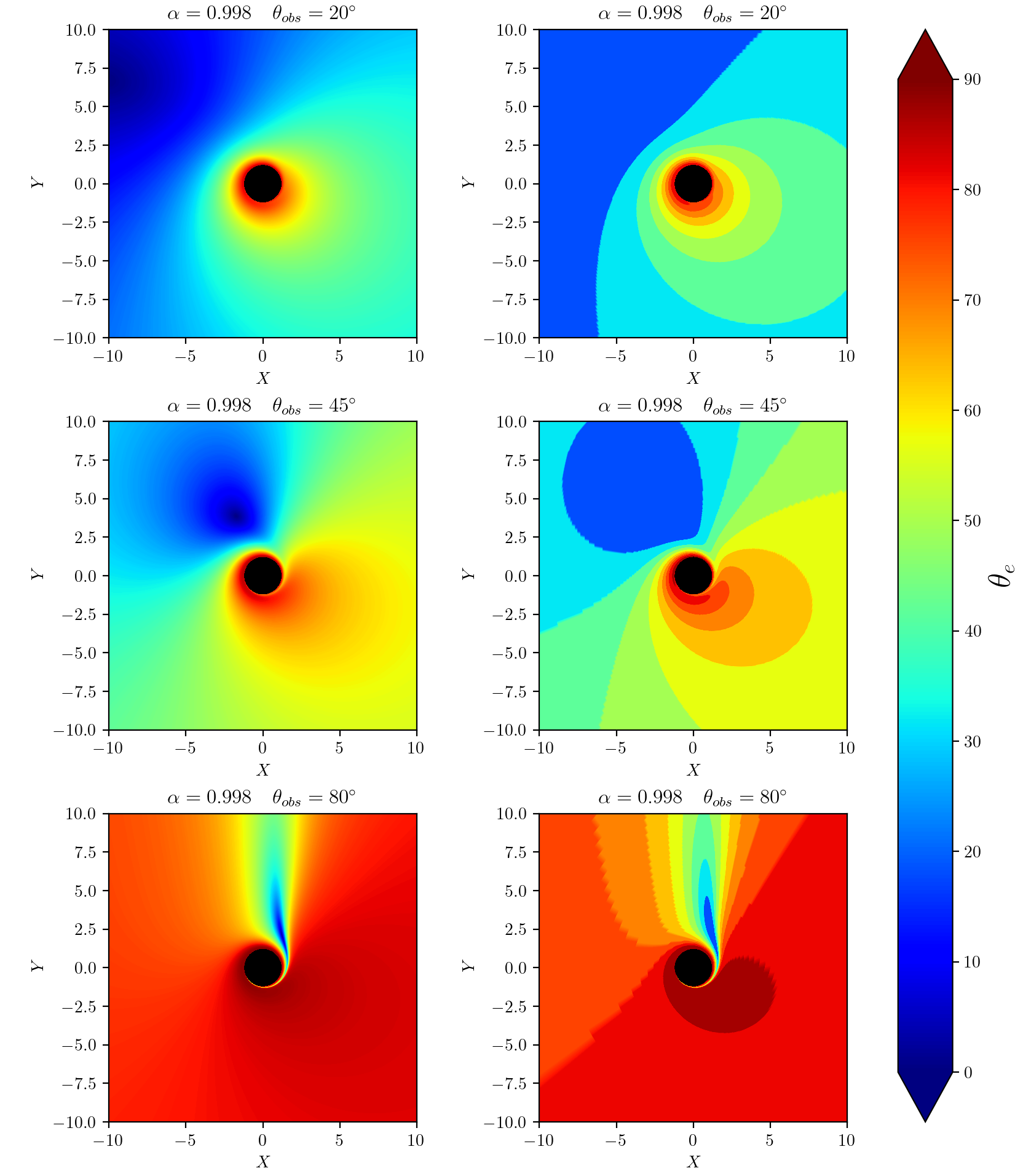}
    \caption{Emission angle map as calculated by the ray-tracing code {\tt blackray} (left panels) and by \texttt{relxillA} (right panels) when the black hole spin parameter is $\alpha = 0.998$ and the inclination angle of the disk with respect to the line of sight of the observer is $\theta_{obs} = 20^\circ$ (top panels), $45^\circ$ (central panels), and $80^\circ$ (bottom panels). The projection of the observer in the $XY$ plane is at $X=0$ and $Y<0$. The black circle at the center of every panel is the region inside the ISCO, where there is no emission.}
    \label{fig:map}
\end{figure*}

Eq.~(\ref{eq-flux}) is obtained after manipulating some variables, but it is exact, without approximations. $I_e$ depends on the exact X-ray spectrum illuminating the disk at the radial coordinate $r_e$, so we should solve numerically the radiative transfer equations for every emission radius $r_e$, which would be unfeasible. Current reflection models like {\tt relxill} have to employ some simplifications in order to be able to generate quickly many relativistic reflection spectra for different values of the model parameters and compare their theoretical predictions with observations. In the specific case of {\tt relxill}, the model calculates the observed spectral flux as
\begin{widetext}
 \begin{eqnarray}\label{eq-flux-approx}
F_{obs} (E_{obs}) &=& \frac{1}{D^2} \int_{R_{\rm in}}^{R_{\rm out}} dr_e \int_0^1 dg^* 
\frac{\pi r_e g^2}{\sqrt{g^* \left(1 - g^*\right)}} 
\left[ f^{(1)} (g^* , r_e , \theta_{obs}) + f^{(2)} (g^* , r_e , \theta_{obs}) \right]
\, \epsilon ( r_e ) \, \langle \bar{I}_e (E_e) \rangle \, ,
\end{eqnarray}  
where $\epsilon$ is the emissivity profile (which depends on how the corona illuminates the accretion disk), $\langle \bar{I}_e (E_e) \rangle$ is the normalized average specific intensity of the radiation
 \begin{eqnarray}\label{eq-flux-approx-2}
\langle \bar{I}_e (E_e) \rangle &=& \frac{1}{C} \sum_{i = 0}^{9}
\int_{R_{\rm in}}^{R_{\rm out}} dr_e \int_0^1 dg^* 
\frac{\pi r_e g^2}{\sqrt{g^* \left(1 - g^*\right)}} 
\left[ f^{(1)} (g^* , r_e , \theta_{obs}) + f^{(2)} (g^* , r_e , \theta_{obs}) \right] \nonumber\\
&& \hspace{4.0cm} \times \epsilon ( r_e ) \, \bar{I}_e (E_e , r_e , \bar{\theta}_e ) \, 
\Theta (\theta_e - \theta_i) \, \Theta (\theta_{i + 1} - \theta_e) \, , \\ \nonumber\\
C &=& \int_{R_{\rm in}}^{R_{\rm out}} dr_e \int_0^1 dg^* 
\frac{\pi r_e g^2}{\sqrt{g^* \left(1 - g^*\right)}} 
\left[ f^{(1)} (g^* , r_e , \theta_{obs}) + f^{(2)} (g^* , r_e , \theta_{obs}) \right]
\, \epsilon ( r_e ) \, , \label{eq-flux-approx-2b}
\end{eqnarray} 
$\Theta$ is the Heaviside step function [$\Theta (x) = 0$ if $x < 0$ and $\Theta (x) = 1$ if $x \ge 0$], $\theta_i = i \, \pi / 20$, $\bar{I}_e (E_e , r_e , \bar{\theta}_e)$ is the normalized specific intensity of the radiation, and $\bar{\theta}_e$ is a value between $\theta_i$ and $\theta_{i + 1}$ [we can use $\bar{\theta}_e = ( \theta_i + \theta_{i + 1} )/2$, but its exact value has no significant impact on the final result]. In the case of {\tt relxilllp} and {\tt relxill\_nk}, the disk is divided into 10 and 50~annuli, respectively, and the observed spectral flux is calculated as
 \begin{eqnarray}
F_{obs} (E_{obs}) &=& \frac{1}{D^2} \sum_{k = 1}^{N} \int_{R_{k}}^{R_{k+1}} dr_e \int_0^1 dg^* 
\frac{\pi r_e g^2}{\sqrt{g^* \left(1 - g^*\right)}} 
\left[ f^{(1)} (g^* , r_e , \theta_{obs}) + f^{(2)} (g^* , r_e , \theta_{obs}) \right]
\, \epsilon ( r_e ) \, \langle \bar{I}_e (E_e) \rangle_k \, , \\ \nonumber\\
\langle \bar{I}_e (E_e) \rangle_k &=& \frac{1}{C_k} \sum_{i = 0}^{9}
\int_{R_k}^{R_{k+1}} dr_e \int_0^1 dg^* 
\frac{\pi r_e g^2}{\sqrt{g^* \left(1 - g^*\right)}} 
\left[ f^{(1)} (g^* , r_e , \theta_{obs}) + f^{(2)} (g^* , r_e , \theta_{obs}) \right] \nonumber\\
&& \hspace{4.0cm} \times \epsilon ( r_e ) \, \bar{I}_e (E_e , r_e , \bar{\theta}_e ) \, 
\Theta (\theta_e - \theta_i) \, \Theta (\theta_{i + 1} - \theta_e) \, , \label{eq-flux-approx-2-2} \\ \nonumber\\
C_k &=& \int_{R_k}^{R_{k+1}} dr_e \int_0^1 dg^* 
\frac{\pi r_e g^2}{\sqrt{g^* \left(1 - g^*\right)}} 
\left[ f^{(1)} (g^* , r_e , \theta_{obs}) + f^{(2)} (g^* , r_e , \theta_{obs}) \right]
\, \epsilon ( r_e ) \, , \label{eq-flux-approx-2-2b}
\end{eqnarray}  
\end{widetext}
where $N$ is the number of annuli ($N = 10$ in {\tt relxilllp} and $N=50$ in {\tt relxill\_nk}) , $R_1 = R_{\rm in}$, $R_{N+1} = R_{\rm out}$, and $R_k < R_{k+1}$ for $k = 1, 2, ... , N$.

The transfer function $f$, the emissivity profile $\epsilon$, and the normalized specific intensity $\bar{I}_e$ are pre-calculated and tabulated into three different FITS files before the data analysis process. During the data analysis process, the model calls the three FITS files, extracts the relevant information, and quickly solves the integral to determine the observed spectral flux $F_{obs} (E_{obs})$.

\section{Improved calculations of the emission angle} \label{sec:2bis}

As we showed in \citet{2025MNRAS.536.2594L}, we cannot fit well high-quality black hole spectra expected from the next generation of X-ray missions with {\tt relxill} and {\tt relxill\_nk}. We found large residuals in the ratios between the simulated data and the best-fit models. A more accurate treatment of the emission angle is necessary.

First, we found that in {\tt relxill} (hereafter, {\tt relxill}~v2.4), the numerators of the fractions inside integrals (\ref{eq-flux-approx-2}) and (\ref{eq-flux-approx-2b}) have $r_e^4$ instead of $r_e$. Similarly, in {\tt relxilllp} and {\tt relxill\_nk} we have $r_e^4$ instead of $r_e$ in Eqs.~(\ref{eq-flux-approx-2-2}) and (\ref{eq-flux-approx-2-2b}). We fixed the expressions of these integrals in a new version of {\tt relxill}, which we called {\tt relxill}~v2.5. As we will show later, this is still not enough to fit high-quality black hole spectra expected from the next generation of X-ray missions.

To improve the calculations of the emission angle, we rewrite Eq.~(\ref{eq-flux-approx}) as follows
\begin{widetext}
 \begin{eqnarray}\label{eq-flux-ai}
F_{obs} (E_{obs}) & = & \frac{1}{D^2} \sum_{i = 0}^{9}  
\int_{R_{\rm in}}^{R_{\rm out}} dr_e  \int_0^1 dg^*
\frac{\pi r_e g^2}{\sqrt{g^* \left(1 - g^*\right)}} 
\left[ f^{(1)} (g^* , r_e , \theta_{obs}) + f^{(2)} (g^* , r_e , \theta_{obs}) \right] \nonumber\\
&& \hspace{4.0cm} \times \epsilon ( r_e ) \, \bar{I}_e (E_e , r_e , \bar{\theta}_e ) \, 
\Theta (\theta_e - \theta_i) \, \Theta (\theta_{i + 1} - \theta_e) \, .
\end{eqnarray}  
\end{widetext}
While this expression of $F_{obs} (E_{obs})$ is more compact, it involves ten convolutions instead of one as in Eq.~(\ref{eq-flux-approx}), and this makes the model somewhat slower than standard  {\tt relxill} based on Eq.~(\ref{eq-flux-approx}). Tab.~\ref{t-speed} reports the average computational time for a reflection spectrum for {\tt relxill}~v2.4, {\tt relxillA}, {\tt relxilllp}~v2.4, and {\tt relxilllpA} on a machine with an Apple M1 chip (8-core~CPU, 8~GB~RAM) using a single core.

\begin{figure*}[t]
    \centering
    \includegraphics[width=1\linewidth,trim=2.7cm 0cm 0.5cm 0cm,clip]{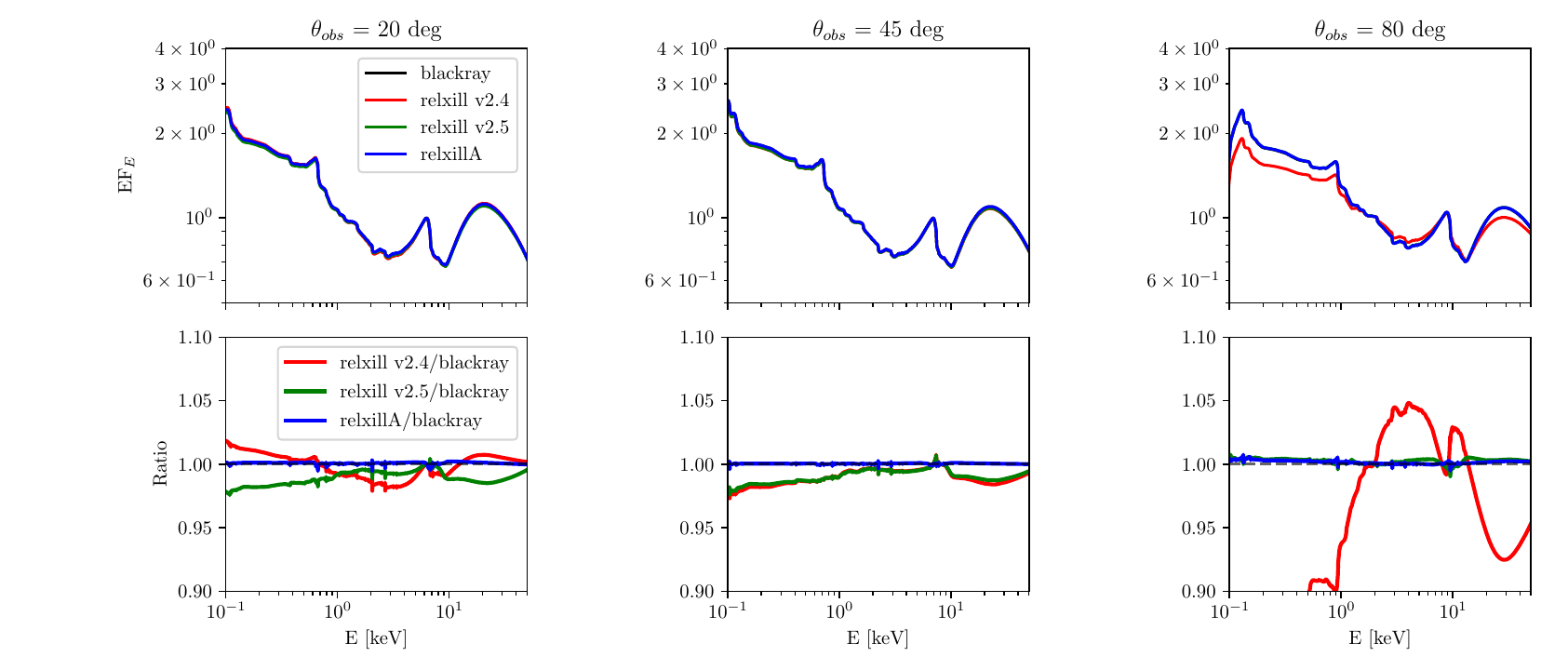}
    \caption{Relativistic reflection spectra as calculated by {\tt blackray}, {\tt relxill}~v2.4, {\tt relxill}~v2.5, and {\tt relxillA} (upper panels) and ratios between the spectra of {\tt blackray} and the spectra of the other models (bottom panels) when the black hole spin parameter is $\alpha=0.998$, and the angle of the observer is $\theta_{obs} = 20^{\circ}$ (left panels), $45^{\circ}$ (central panels), and $80^{\circ}$ (right panels). Spectra are normalized at the peak of the iron line to facilitate their comparison. See the text for more details. \label{fig:spectrum}}
    \vspace{0.5cm}
\end{figure*}

\begin{table}[h!]
\centering
\begin{tabular}{lcc}
\hline\hline
Model  & \hspace{0.3cm} N\_ENER\_CONV \hspace{0.3cm} & $\tau$~(ms) \\
{\tt relxill}~v2.4 & 4096 & 14 \\
{\tt relxillA} & 4096 & 17 \\
{\tt relxilllp}~v2.4 & 4096 & 28 \\
{\tt relxilllpA} & 4096 & 36 \\ 
\hline
{\tt relxill}~v2.4 & 32768 & 70 \\
{\tt relxillA} & 32768 & 90 \\
{\tt relxilllp}~v2.4 & 32768 & 95 \\
{\tt relxilllpA} & 32768 & 165 \\
\hline\hline
\end{tabular}\\
\vspace{0.3cm}
\caption{Average computational time $\tau$ of a reflection spectrum on a machine with an Apple M1 chip (8-core CPU, 8 GB RAM), using a single core, for {\tt relxill}~v2.4, {\tt relxillA}, {\tt relxilllp}~v2.4, and {\tt relxilllpA} and two different values of N\_ENER\_CONV (4096 is the default value in the {\tt relxill} package and 32768 is the value used in \citet{2025MNRAS.536.2594L} and in Section~\ref{sec:sim} to fit high-quality \textit{NewAthena}/X-IFU+LAD spectra); see \citet{2025MNRAS.536.2594L} for more details on this parameter. We averaged the computational time over different values of the black hole spin, inclination angle of the disk, and iron abundance. \label{t-speed} }
\end{table}

The left panels in Fig.~\ref{fig:map} shows the values of the emission angle $\theta_e$ over the disk as calculated by the ray-tracing code {\tt blackray} \citep{abdikamalov_2024_10673859}. This model is accurate, but too slow to analyze observations. We assume that the spacetime is described by the Kerr solution with black hole spin parameter $\alpha=0.998$, the inner edge of the accretion disk is at the innermost stable circular orbit (ISCO) of the spacetime, there is no emission inside the ISCO, and the inclination angle of the disk with respect to the line of sight of the distant observer is $\theta_{obs} = 20^\circ$ (top panel), $45^\circ$ (central panel), and $80^\circ$ (bottom panel). The black circle at the center of every panel is the region inside the ISCO, where there is no emission. The right panels in Fig.~\ref{fig:map} shows the values of the emission angle $\theta_e$ as calculated by our new model {\tt relxillA}. The 10~zones with constant emission angle have different colors and we can see that the shapes of the zones depend significantly on the inclination angle of the disk $\theta_{obs}$.

At this point, we have four reflection models: the accurate but slow {\tt blackray}, {\tt relxill}~v2.4 with the incorrect $r_e^4$ term, {\tt relxill}~v2.5 with the correct $r_e$ term, and the new model {\tt relxillA} obtained by replacing the calculation of the observed spectral flux in Eqs.~(\ref{eq-flux-approx})-(\ref{eq-flux-approx-2b}) with the calculation in Eq.~(\ref{eq-flux-ai}).

\begin{figure*}
    \centering    
    \includegraphics[width=1\linewidth,trim=2.7cm 0cm 0.5cm 0cm,clip]{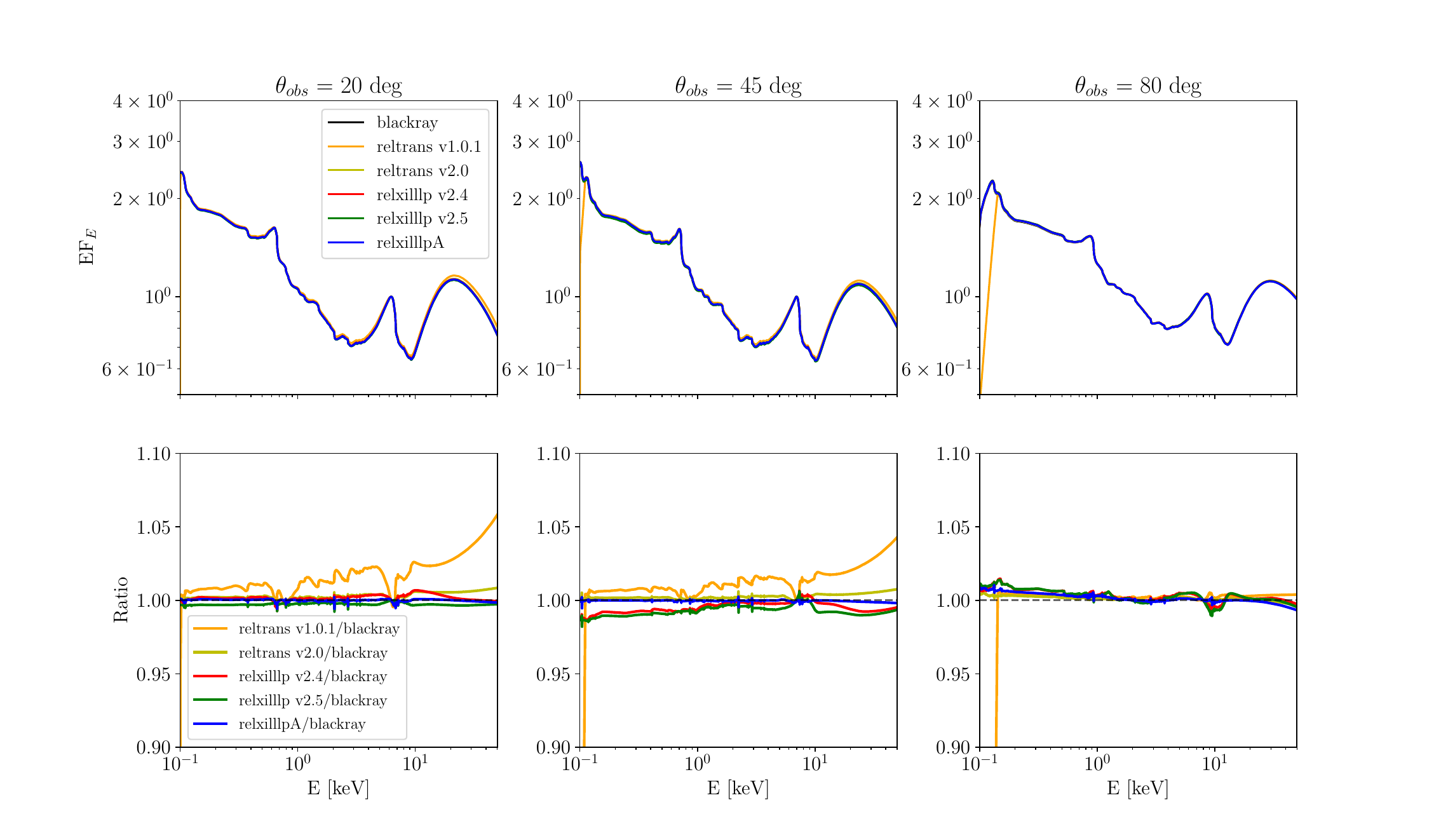}
    \caption{Relativistic reflection spectra as calculated by {\tt blackray}, {\tt reltrans}~v1.0.1, {\tt reltrans}~v2.0, {\tt relxilllp}~v2.4, {\tt relxilllp}~v2.5, and {\tt relxilllpA} (upper panels) and ratios between the spectra of {\tt blackray} and the spectra of the other models (bottom panels) when the black hole spin parameter is $\alpha=0.998$, and the angle of the observer is $\theta_{obs} = 20^{\circ}$ (left panels), $45^{\circ}$ (central panels), and $80^{\circ}$ (right panels). Spectra are normalized at the peak of the iron line to facilitate their comparison. See the text for more details. \label{fig:spectrumlp}}    
\end{figure*}

\begin{figure}
    \centering
    \includegraphics[width=1\linewidth,trim=0cm 0cm 0cm 0cm,clip]{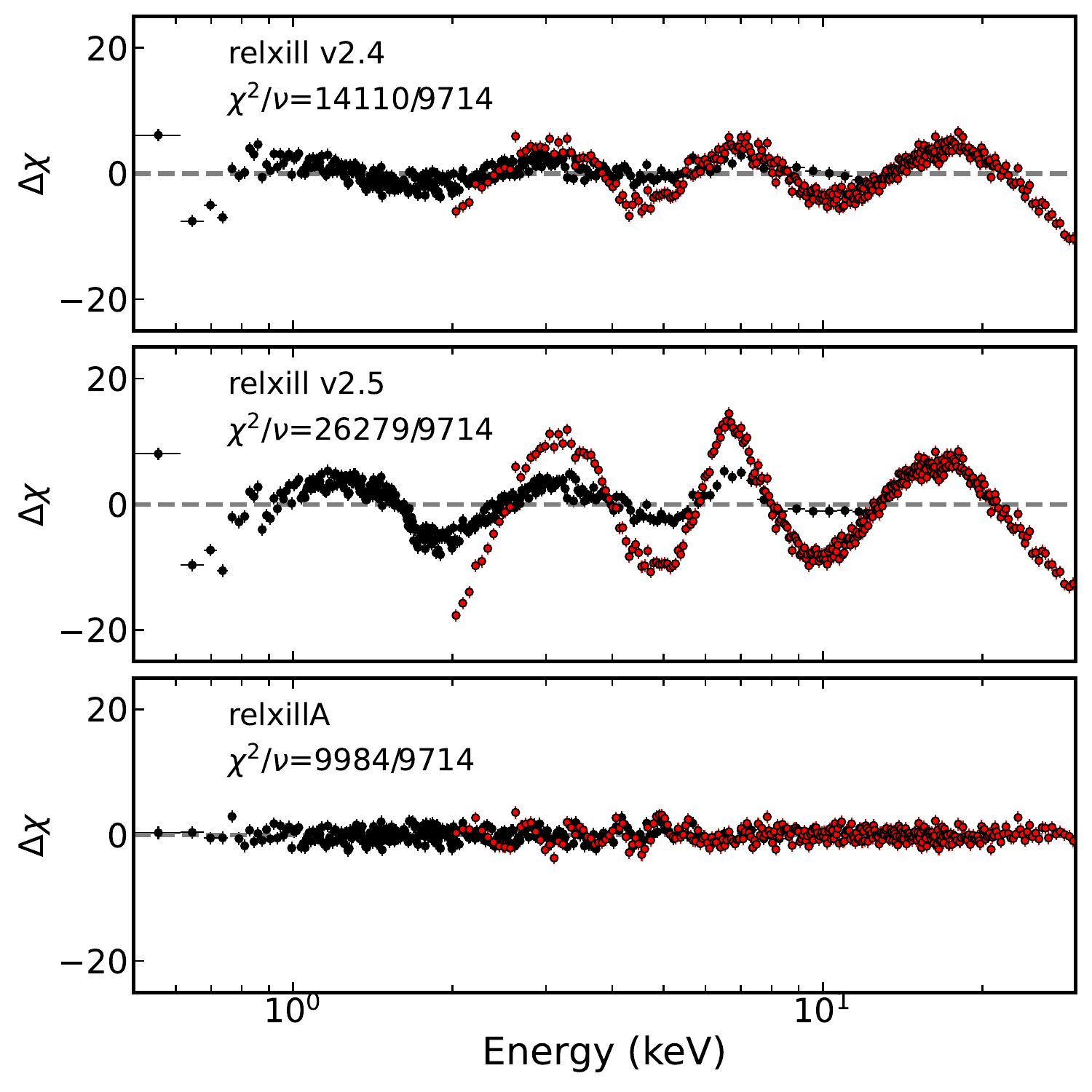}
    \caption{Residuals of the best-fit models of {\tt relxill}~v2.4, {\tt relxill}~v2.5, and {\tt relxillA} of a simulated \textit{NewAthena}/X-IFU+LAD spectrum of a bright black hole X-ray binary with spin parameter $\alpha = 0.998$ and inclination angle of the disk $\theta_{obs} = 15^\circ$. The black color is used for the \textit{NewAthena}/X-IFU data and the red color is for the LAD data.}
    \label{fig:sim}
\end{figure}

Fig.~\ref{fig:spectrum} compares the relativistic reflection spectra calculated by {\tt blackray}, {\tt relxill}~v2.4, {\tt relxill}~v2.5, and {\tt relxillA}. We consider the specific case of a Kerr spacetime with black hole spin parameter $\alpha = 0.998$ and we show the predictions for $\theta_{obs} = 20^\circ$ (left panels), $45^\circ$ (central panels), and $80^\circ$ (right panels). We assume the following values of the {\tt relxill} parameters: photon index $\Gamma = 2$, emissivity index $q = 3$, ionization parameter $\log\xi = 3.1$ ($\xi$ in units erg~cm~s$^{-1}$), iron abundance $A_{\rm Fe} = 1$, high-energy cutoff $E_{\rm cut} = 300$~keV, and reflection fraction $R_{\rm f} = -1$ (pure reflection). The top panels show the relativistic reflection spectra of the four models. The bottom panels show the differences between {\tt blackray} and the other three models. All spectra are normalized at the peak of the iron line to facilitate the comparison.

For low and moderate values of the inclination angle of the disk ($\theta_{obs} = 20^\circ$ and $45^\circ$), the performance of {\tt relxill}~v2.4 and {\tt relxill}~v2.5 is quite similar and their relative difference with respect to the spectrum calculated by {\tt blackray} is within 2\% over the whole X-ray band. When the inclination angle is very high ($\theta_{obs} = 80^\circ$), the error in the calculation of the normalized specific intensity of the radiation $\langle \bar{I}_e (E_e) \rangle$ in {\tt relxill}~v2.4 produces deviations at the level of 8\% at the Compton hump and more than 10\% below 1~keV. On the contrary, the difference between {\tt relxill}~v2.5 and {\tt blackray} is always below 1\% for $\theta_{obs} = 80^\circ$. The fact that {\tt relxill}~v2.5 is more accurate for $\theta_{obs} = 80^\circ$ can be explained by looking at Fig.~\ref{fig:map}, where we see that for $\theta_{obs} = 80^\circ$ there is only a narrow region of the disk emitting with a very different emission angle, so the approximation in Eqs.~(\ref{eq-flux-approx})-(\ref{eq-flux-approx-2b}) works better. Last, the differences between {\tt relxillA} and {\tt blackray} are very small and difficult to see in these plots.

At this point, it can be useful to check even the accuracy of {\tt reltrans}~\citep{2019MNRAS.488..324I}, which is another publicly available reflection model. Since {\tt reltrans} assumes a lamppost setup, we have to run our ray-tracing code {\tt blackray} with a lamppost coronal model. Fig.~\ref{fig:spectrumlp} shows the relativistic reflection spectra calculated by {\tt blackray}, {\tt reltrans}~v1.0.1~\citep{2019MNRAS.488..324I}, {\tt reltrans}~v2.0~\citep{2021MNRAS.507...55M}, {\tt relxilllp}~v2.4, {\tt relxilllp}~v2.5, and {\tt relxilllpA}. Again, we consider the specific case of a Kerr spacetime with black hole spin parameter $\alpha = 0.998$ and we show the predictions for $\theta_{obs} = 20^\circ$ (left panels), $45^\circ$ (central panels), and $80^\circ$ (right panels). The values of the model parameters are the same as in Fig.~\ref{fig:spectrum}, but we assume a lamppost corona and we set the coronal height $h = 5$~$R_{\rm g}$ ($R_{\rm{g}} = G_{\rm N} M /c^2$ is the gravitational radius of the black hole). In {\tt reltrans}~v1.0.1, the convolution kernel is divided into three dimensions: emission angle ($\mu=\cos\theta_e$), high-energy cutoff of the coronal spectrum ($E_{\rm cut}$), and disk ionization. We set 10~zones in $\mu$ and 10~zones in $E_{\rm cut}$, with the ionization parameter and the electron density fixed to single constant values across the entire disk. {\tt reltrans}~v2.0 simplifies this structure into two dimensions: emission angle and radial zones. We set 10~zones for the emission angle $\mu$ and 10~radial zones, each of them with a different $E_{\rm cut}$ value. The electron density and the ionization parameter are fixed and remain constant across all radial zones. The top panels show the relativistic reflection spectra of the six models. The bottom panels show the differences between {\tt blackray} and the other five models. All spectra are normalized at the peak of the iron line to facilitate the comparison.

For low and moderate values of the inclination angle of the disk ($\theta_{obs} = 20^\circ$ and $45^\circ$), the performance of {\tt reltrans}~v1.0.1 is worse than the other models because the choice of dividing the disk into $E_{\rm cut}$-zones does not lead to an accurate calculation of the reflection spectra. This is fixed in {\tt reltrans}~v2.0, where the disk is divided into annuli and every annulus has its $E_{\rm cut}$ value. For moderate and high inclination angles of the disk ($\theta_{obs} = 45^\circ$ and $80^\circ$), {\tt reltrans}~v2.0 turns out to be more accurate than {\tt relxilllp}~v2.4 and {\tt relxilllp}~v2.5 below 10~keV, while its deviations from the ray-tracing code {\tt blackray} are comparable to those of {\tt relxilllp}~v2.4 and {\tt relxilllp}~v2.5 above 10~keV. For low inclination angles ($\theta_{obs} = 20^\circ$), the accuracy of {\tt reltrans}~v2.0 is very similar to that of {\tt relxilllp}~v2.4 and {\tt relxilllp}~v2.5.

\section{Simulations with NewAthena/X-IFU+LAD} \label{sec:sim}

In \citet{2025MNRAS.536.2594L}, we simulated a set of observations with the X-IFU instrument onboard \textit{NewAthena}~\citep{2013arXiv1306.2307N} and the LAD instrument~\citep{2016SPIE.9905E..1QZ}\footnote{The Large Area Detector (LAD) is a high throughput instrument that was proposed to be part of the \textit{eXTP} payload~\citep{2016SPIE.9905E..1QZ}. Its nominal energy range is 2-30~keV. At 6~keV, the effective area is 3.4~m$^2$ and the energy resolution is better than 250~eV.}, which may also fly in future. We assumed the observation of very bright black hole X-ray binaries (we set the source flux to $1 \times 10^{-7}$~erg~s$^{-1}$~cm$^{-2}$ in the 2-10~keV band) and an exposure time of 100~ks for each instrument. We showed that the approximations implemented in {\tt relxill} and {\tt relxill\_nk} are unsuitable to fit high-quality full reflection spectra, which require instead reflection models with more accurate calculations of the emission angle. Both models were able to recover the correct input parameters, but the ratios between the simulated data and the best-fit models presented large residuals.

Fig.~\ref{fig:sim} shows that {\tt relxillA} solves this problem. We use a simulated observation from \citet{2025MNRAS.536.2594L}. The total spectrum has a power law component to describe the coronal spectrum and a relativistic reflection component to describe the reflection spectrum from the disk. For the input parameters, we assume the following values: black hole spin parameter $\alpha = 0.998$, inclination angle of the disk $\theta_{obs} = 15^\circ$, photon index $\Gamma = 1.7$, high-energy cutoff of the power law component $E_{\rm cut} = 300$~keV, emissivity profile of the disk described by a power law with emissivity index $q = 5$, ionization of the disk $\log\xi = 1.0$, and iron abundance $A_{\rm Fe} = 1$. We fit the simulated data with {\tt relxill}~v2.4, {\tt relxill}~v2.5, and {\tt relxillA} and we show the residuals in Fig.~\ref{fig:sim}. {\tt relxill}~v2.5 does not do better than {\tt relxill}~v2.4, while there are no residuals in the fit of {\tt relxillA}.

We repeated some simulations by replacing \textit{NewAthena}/X-IFU with \textit{XRISM}/Resolve and LAD with \textit{NuSTAR} to figure out if \textit{XRISM} can reveal the limitations of the older models and distinguish the improvements offered by {\tt relxillA}. We found that all models can fit the data well without residuals: the large effective areas of \textit{NewAthena}/X-IFU and LAD are crucial to appreciate the improvements offered by {\tt relxillA}.

\section{An example fit: EXO~1846--031} \label{sec:3}

EXO~1846--031 is a black hole X-ray binary first discovered by \textit{EXOSAT} in 1985 \citep{1985IAUC.4051....1P}. It experienced a new outburst in 2019 \citep{2019ATel12968....1N}, and was observed by \textit{NuSTAR} during the hard-to-intermediate state transition (on 2019-08-03, ObsID 90501334002) \citep{2019ATel13012....1M}. Previous studies have revealed that the spectrum of this observation shows strong reflection features \citep{2019ATel13012....1M}, making it ideal for the study of new reflection models \citep[e.g., ][]{Askar2021_relxillnk,Ashutosh2021_Relxillnk,Songcheng2024}. Through reflection modeling, the source has also been found to have a high spin close to the maximum and a high disk inclination angle of about $73^{\circ}$ \citep{Draghis2020_spin_incl}. As shown in Fig.~\ref{fig:spectrum}, {\tt relxill}~v2.4 presents the largest deviations from {\tt blackray} when the source is observed at a high inclination angle. Therefore, EXO 1846--031 is a particularly suitable source to see possible differences between the parameter estimates of {\tt relxill}~v2.4 and the more accurate model {\tt relxillA}.

We reduce the data with \texttt{NuSTARDAS}\footnote{https://heasarc.gsfc.nasa.gov/docs/nustar/analysis/} and \texttt{CALDB 20250317} \citep{2022JATIS...8c4003M}. We select a circular region with a radius of $180''$ on both Focal Plane Module A (FPMA) and Focal Plane Module B (FPMB) detectors to extract the source spectra. A circular region of comparable size far away from the source region is selected to extract the background spectra. Afterwards, we use \texttt{nuproducts} to generate the source and background spectra. We group the spectra with the optimal binning algorithm by using the \texttt{ftgrouppha} task\footnote{https://heasarc.gsfc.nasa.gov/lheasoft/help/ftgrouppha.html}. The bright source treatment is applied by modifying the \texttt{statusexpr} keyword in \texttt{nupipeline}\footnote{https://heasarc.gsfc.nasa.gov/docs/nustar/nustar\_faq.html\#bright}. Spectral fittings are conducted with \texttt{XSPEC} v12.13.0 \citep{Arnaud1996}. We implement the element abundances of \citet{Wilms_2000} and cross-sections of \citet{Verner1996}. The $\chi^2$ statistics is used to find the best fit values and uncertainties (at the 90\% of confidence level) of the parameters.

We fit the data with three models 

\vspace{0.2cm}

Model~1:

\texttt{constant*tbabs*(diskbb+cutoffpl+relxill~\rm{v2.4})}

\vspace{0.2cm}

Model~2:

\texttt{constant*tbabs*(diskbb+cutoffpl+relxill~\rm{v2.5})}

\vspace{0.2cm}

Model~3:

\texttt{constant*tbabs*(diskbb+cutoffpl+relxillA)}

\vspace{0.2cm}

\noindent There are three components: the thermal emission from the disk \citep[\texttt{diskbb},][]{Mitsuda1984_diskbb}, the Comptonized corona emission approximated by a power-law with a high-energy cutoff (\texttt{cutoffpl}), and the reflection component (\texttt{relxill}~v2.4, \texttt{relxill}~v2.5, and \texttt{relxillA}, respectively in the first, second, and third model). \texttt{tbabs} is used to account for the interstellar absorption \citep{Wilms_2000}. We also multiply the model with a \texttt{constant} for cross normalization between the FPMA and FPMB data. We link the high-energy cutoff ($E_{\rm cut}$) and the photon index ($\Gamma$) in the Comptonization component and reflection component together. The inner radius of the accretion disk is fixed at the ISCO radius\footnote{This assumption can be justified by the fact that the source was in an intermediate state at high luminosity. However, leaving the inner radius of the disk free in the fit does not change the results: the data require in any case that the inner edge of the disk is as close as possible to the black hole, so it should be close to the ISCO radius of a very fast-rotating black hole.} and the outer edge is fixed at 1000~$R_{\rm{g}}$. The reflection fraction is frozen at $-1$ because the corona emission is already described by \texttt{cutoffpl}.

\begin{figure*}\label{fit_plot}
    \centering
    \includegraphics[width=0.32\linewidth]{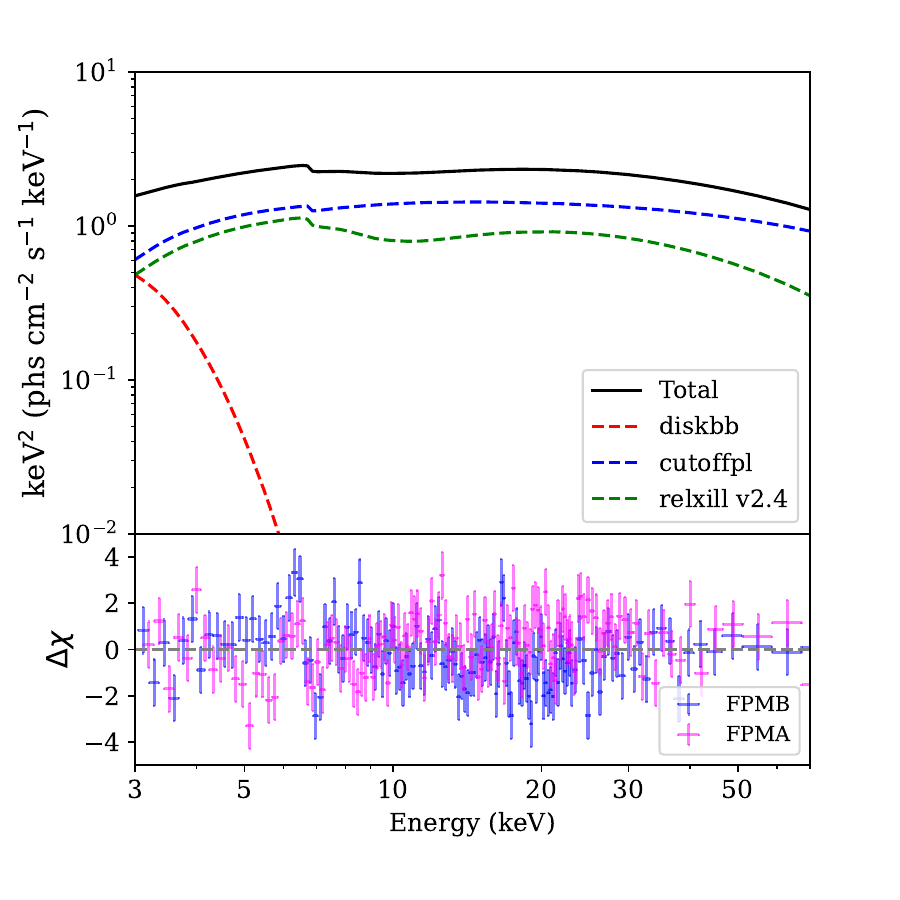}
    \includegraphics[width=0.32\linewidth]{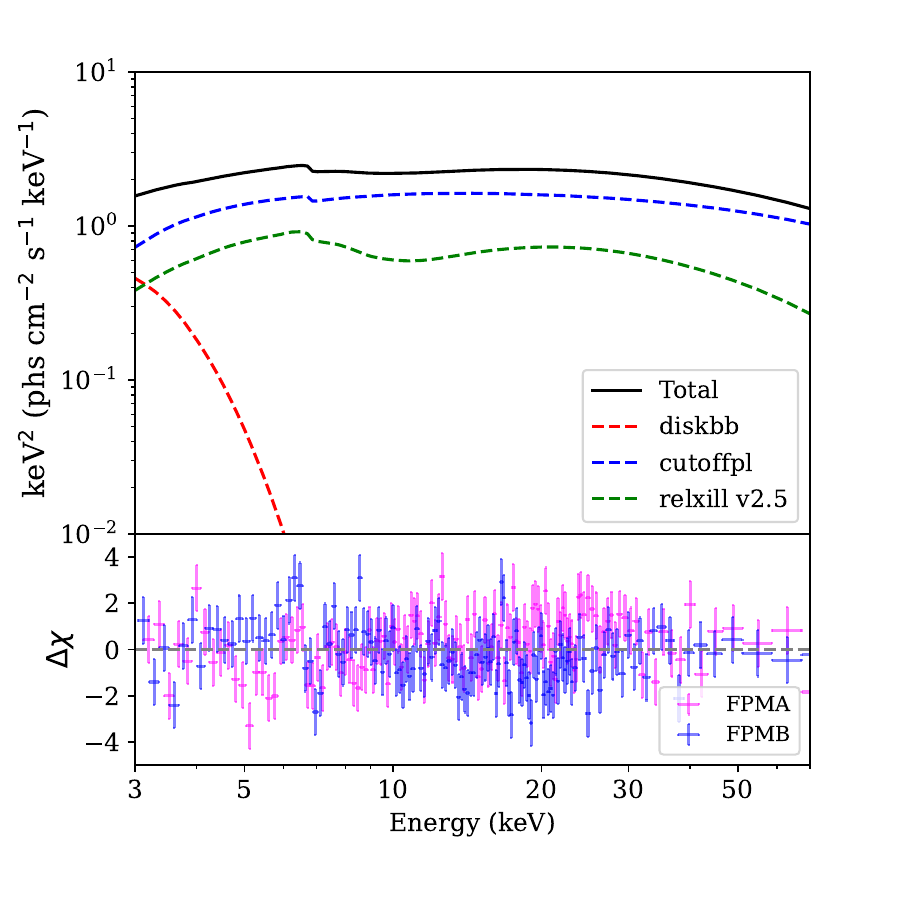}
    \includegraphics[width=0.32\linewidth]{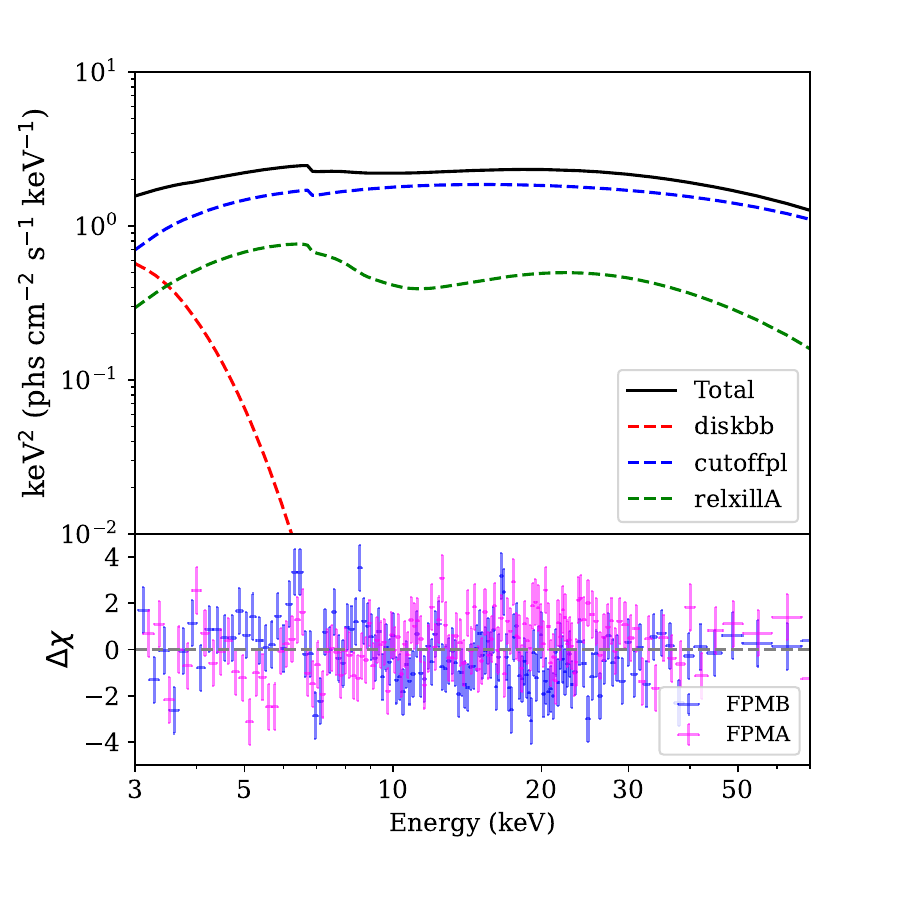} \\
    \vspace{-0.4cm}
    \caption{Best-fit model and fitting residuals for Model~1 (left panel), Model~2 (central panel), and Model~3 (right panel). In the upper quadrants, we show the total model and different model components in different colors. In the lower quadrants, we show the fitting residuals of FPMA and FPMB data.}
    \vspace{0.5cm}
\end{figure*}

The best-fit parameter values of the three models are shown in Tab.~\ref{fit_table}. The best-fit models and residuals are shown in Fig.~\ref{fit_plot}.

The best-fit of Model~1 has $\chi^2 = 635.66/526=1.21$ and the estimate of the model parameters is consistent with previous studies \citep{Draghis2020_spin_incl,Askar2021_relxillnk,Ashutosh2021_Relxillnk,Songcheng2024}: the spin parameter $\alpha$ is extremely high, the disk inclination angle is over $70^{\circ}$, and the emissivity profile can be described well by a broken power law with a very steep inner emissivity profile and an almost flat outer emissivity profile. The fit can be slightly improved by employing a reflection model with a non-vanishing ionization gradient, but this does not affect the estimate of the parameters of the model~\citep{Songcheng2024}. The emissivity profile of this spectrum was studied in details in \citet{Songcheng2024}, where it is shown that alternative solutions yield significantly poorer best-fit results. Similar emissivity profiles have been reported even from the studies of other sources, like GS~1354--645~\citep{2018ApJ...865..134X} and GRS~1915+105~\citep{2019ApJ...884..147Z}, and can be interpreted as generated by a corona with a ring-like axisymmetry geometry located just above the accretion disk~\citep{2003MNRAS.344L..22M,2011MNRAS.414.1269W,2022ApJ...925...51R}\footnote{The emissivity profile from such a coronal geometry would be approximated better by a twice broken power law, with a very steep emissivity profile in the inner region, a flat or almost flat emissivity profile in the intermediate region, and an emissivity profile falling off approximately as $1/r^3$ in the outer region~\citep{2003MNRAS.344L..22M,2011MNRAS.414.1269W,2022ApJ...925...51R}.}.

In the case of Model~2 and Model~3, the quality of the fits is very similar ($\chi^2 = 635.29/526=1.21$ and $639.80/526=1.22$, respectively). For Model~2, the estimate of all parameters is consistent with the results of Model~1. For Model~3, the estimate of most parameters is consistent with the results of Model~1 and Model~2: minor discrepancies are in the estimate of the inner emissivity index $q_{\rm in}$ and the black hole spin parameter $\alpha$ (which are not stuck to the upper bounds as in Model~1 and Model~2) and a somewhat lower value of the inclination angle of the disk.

\vspace{1.0cm}

\begin{table*}[h!]
\centering
\begin{tabular}{lccc}
\hline\hline
& Model~1 & Model~2 & Model~3 \\
& ({\tt relxill}~v2.4) & ({\tt relxill}~v2.5) & ({\tt relxillA}) \\
Parameters &&& \\
\hline\hline
\texttt{tbabs} & & &\\
$n_{\rm{H}}$ [$10^{22}$~cm$^{-2}$] & $10.2^{+0.4}_{-0.6}$ & $9.9^{+0.3}_{-0.4}$ & $10.9^{+0.1}_{-1.5}$\\
\hline
\texttt{diskbb} & & &\\
$T_{\rm{in}}$ [keV] & $0.47^{+0.01}_{-0.01}$& $0.46^{+0.01}_{-0.01}$ & $0.487^{+0.001}_{-0.001}$\\
norm &$21000^{+2000}_{-4000}$& $20000^{+2000}_{-3000}$ & $17900^{+300}_{-600}$\\
\hline
\texttt{cutoffpl}&&&\\
$\Gamma$ &$1.92^{+0.01}_{-0.02}$& $1.93^{+0.04}_{-0.04}$&$1.834^{+0.001}_{-0.001}$\\
$E_{\rm cut}$ [keV] &$82^{+3}_{-9}$& $86^{+3}_{-6}$&$68.61^{+0.75}_{-0.03}$\\
norm &$1.7^{+0.1}_{-0.1}$& $1.85^{+0.05}_{-0.3}$&$1.56^{+0.003}_{-0.006}$\\
\hline
\texttt{relxill/relxillA}& & &\\
$q_{\rm in}$ & $10.0_{-0.3}$& $10.0_{-0.5}$& $7.6^{+0.4}_{-0.1}$\\
$q_{\rm out}$ & $0.4^{+0.6}_{-0.4}$& $0.3_{-0.2}^{+0.7}$& $1.2^{+0.1}_{-0.3}$\\
$R_{\rm{br}}$ $[R_{\rm{g}}]$ & $6^{+1}_{-1}$& $7^{+1}$& $8.2^{+0.9}_{-0.6}$\\
$\alpha$ & $0.998_{-0.001}$& $0.998_{-0.001}$& $0.9941^{+0.0004}_{-0.0004}$\\
$\theta_{obs}$ [deg] & $74.6^{+0.2}_{-0.8}$& $75^{+1}_{-2}$& $68.8^{+0.3}_{-0.2}$\\
$\log{\xi}$ [erg~cm~s$^{-1}$] & $3.2^{+0.05}_{-0.07}$ & $3.25^{+0.04}_{-0.03}$& $3.7^{+0.1}_{-0.2}$\\
$A_{\rm{Fe}}$ & $3.3^{+0.8}_{-0.3}$ & $3.3^{+0.8}_{-0.3}$& $3.9^{+1.2}_{-0.5}$ \\
norm & $0.063^{+0.004}_{-0.003}$ & $0.041^{+0.009}_{-0.003}$& $0.021^{+0.082}_{-0.001}$\\
\hline
C$_{\rm{FPMB}}$ &$0.956^{+0.001}_{-0.001}$ & $0.956^{+0.001}_{-0.001}$ &$0.956^{+0.001}_{-0.001}$\\
$\chi^2$/d.o.f & 635.66/526=1.21 & 635.29/526=1.21 & 639.80/526=1.22\\
\hline\hline
\end{tabular}\\
\vspace{0.3cm}
\caption{The best-fit values of the \textit{NuSTAR} observation of EXO 1846--031 with Model~1, Model~2, and Model~3. The uncertainties are calculated at the 90\% of confidence level. When there is no lower/upper uncertainty, the best-fit value is stuck at the lower/upper boundary of the range in which the parameter can vary. \label{fit_table} }
\end{table*}

\section{Summary and conclusions} \label{sec:4}

Current reflection models make a number of approximations in order to be able generate quickly many relativistic reflection spectra and scan the parameter space to find the best-fit model for the spectrum of an observation. One of these approximations is to employ an average non-relativistic reflection spectrum over the whole disk (or over an annulus of the disk when the disk is divided in a number of annuli). As shown in \citet{2025MNRAS.536.2594L}, if we assume such an approximation in the calculation of relativistic reflection spectra, we cannot fit high-quality black hole spectra expected from the next generation of X-ray missions. We also found a bug in {\tt relxill}~v2.4 (see Section~\ref{sec:2bis}). In this work, we have presented a reflection model with improved calculations of the emission angle that solves this problem. Fig.~\ref{fig:sim} shows that our new model can fit well the simulated observations of \citet{2025MNRAS.536.2594L}.

In Section~\ref{sec:3}, we used our new model to fit a high-quality \textit{NuSTAR} spectrum of the black hole binary EXO~1846--031. While the source and its spectrum present promising properties to maximize the effect of light bending, we do not find any significant difference between the fits with {\tt relxill}~v2.4, {\tt relxill}~v2.5, and the new model. Such a result suggests that even past analyses with {\tt relxill}~v2.4 and previous versions are accurate for the available X-ray data of black hole binaries, while we need a reflection model with improved calculations of the emission angle only in the case of higher quality data.

Last, we want to stress that the new model does not introduce any new parameter: {\tt relxillA} simply improves the calculation of the emission angle and predicts slightly different spectra with respect to older models. Therefore it does not introduce any new parameter degeneracy. In the case of low signal-to-noise datasets, we cannot distinguish/appreciate the improvements offered by {\tt relxillA} with respect to the older models (this is the case of the high-quality \textit{NuSTAR} spectrum analyzed in Section~\ref{sec:3}). In the case of high signal-to-noise datasets, we can fit the data well and there are no residuals as with the older models (this is the case of the simulation analyzed in Section~\ref{sec:sim}).

\begin{acknowledgments}
This work was supported by the National Natural Science Foundation of China (NSFC), Grant Nos.~12250610185 and 12261131497, and the Natural Science Foundation of Shanghai, Grant No.~22ZR1403400. 
TD acknowledges support from the DFG research unit FOR~5195 (Project No.~443220636, Grant No.~WI~1860/20-1).
GM acknowledges financial support from the European Union’s Horizon Europe research and innovation programme under the Marie Sk\l{}odowska-Curie grant agreement No.~101107057.
\end{acknowledgments}

\bibliography{references}{}
\bibliographystyle{aasjournal}

\end{document}